# An optimal mode selection algorithm for scalable video coding

L. Balaji

Department of ECE,
Vel Tech Rangarajan Dr. Sagunthala R&D Institute of Science and Technology,
Avadi, Chennai 600062, India
Email: maildhanabal@gmail.com

K.K. Thyagharajan

Department of ECE, RMD Engineering College, Chennai 601206, India
Email: kkthyagharajan@yahoo.com

C. Raja

Department of ECE, Vignan's Foundation for Science Technology and Research,
Vadlamudi, Guntur Dt., Andhra Pradesh, 522213, India
Email: rajachandru82@yahoo.co.in

A. Dhanalakshmi

Department of EIE, Panimalar Engineering College, Chennai 600123, India
Email: dhanalakshmi248@gmail.com

**Abstract:** Scalable video coding (SVC) is extended from its predecessor advanced video coding (AVC) because of its flexible transmission to all type of gadgets. However, SVC is more flexible and scalable than AVC, but it is more complex in determining the computations than AVC. The traditional full search method in the standard H.264 SVC consumes more encoding time for computation. This complexity in computation need to be reduced and many fast mode decision (FMD) algorithms were developed, but many fail to balance in all the three measures such as peak signal to noise ratio (PSNR), encoding time and bit rate. In this paper, the proposed optimal mode selection algorithm based on the orientation of pixels achieves better time saving, good PSNR and coding efficiency. The proposed algorithm is compared with the standard H.264 JSVM reference software and found to be 57.44% time saving, 0.43 dB increments in PSNR and 0.23% compression in bit rate.

## 1   Introduction

The H.264 scalable video coding (H.264/SVC) facilitates to create a bit stream by encoding a video signal once and allows extracting various sub streams with different bit rates and resolutions from the same stream. The sub stream can be scaled over and over until it has smallest description called base layer (BL) stream. The BL is the lowest layer and it has the lowest resolution, the lowest frame rate and the lowest quality of content (Schwarz et al., 2007). This layer is backward compatible with H.264 advanced video coding (H.264/AVC) and it can be decoded independently. The video data can be enhanced to a higher resolution or a higher frame rate by adding additional bit streams of higher layers called the enhancement layers (ELs). Determining or predicting the EL data from the lower layer (reference layer) of the same instant of time is known as the inter-layer prediction process. This makes the video to have various spatial resolutions and this called as spatial scalability. The scalable video coding (SVC) supports three scalabilities viz. spatial, temporal and quality scalabilities. In a multi-layer spatial scalability approach, each layer has a distinct size and it is referred by an identifier D called dependency identifier (Segall and Sullivan, 2007). In SVC two inter-layer prediction concepts (Schwarz et al., 2004) have been introduced:

1. MB mode prediction and associated motion parameters estimation
2. Residual signal prediction

The offset between the location of the current (target) MB and the best-matched block in the reference MB is called motion vector (MV). The offset between the location of the current block and the predicted block in the reference frame is called predicted motion vector (PMV). Spatial correlation is used to predict PMV. The difference between PMV and MV of the MB is called motion vector difference (MVD). If the PMV is predicted accurately a small MVD will be obtained.

The temporal scalability in H.264/SVC is achieved by determining the content of a frame at higher temporal level from the frames at lower levels. The number of frames to be played with an interval of time (temporal resolution) can be changed by applying temporal scalability. The I or P frame in a video is referred to as a key picture or a key frame. This type of frame will be the first frame of a Group of Picture (GOP). A GOP is a number of pictures between two key frames and it will contain 2/4/8/16/32 frames depending on the number of layers used for temporal scalability. The number of frames in a GOP may be given as $2^N$ where N is the number of layers and its value is 0 for the BL (Lee and Kim, 2012).

Figure 1 shows the temporal scalability levels in a spatial base and a spatial EL. Each spatial layer either base or enhancement includes the corresponding temporal scalability levels as chosen by the coder. For a GOP size of 8, there will be 4 temporal scalability levels (TSL0, TSL1, TSL2 and TSL3) in each spatial base and spatial ELs where N range from 0, 1, 2 and 3 or N = 3. The key frame (I or P) will always be at TSL0 that is used to code the upper level (TSL1, TSL2 and TSL3) temporal frames. Figure 1 show that I or P frames alone can be used at TSL0 and B frames are used at TSL1, TSL2 and TSL3 in both spatial base and spatial ELs. Since four temporal levels are included, hence the total number of frames in spatial base and spatial enhancements layers is 8 ($2^3 = 8$).

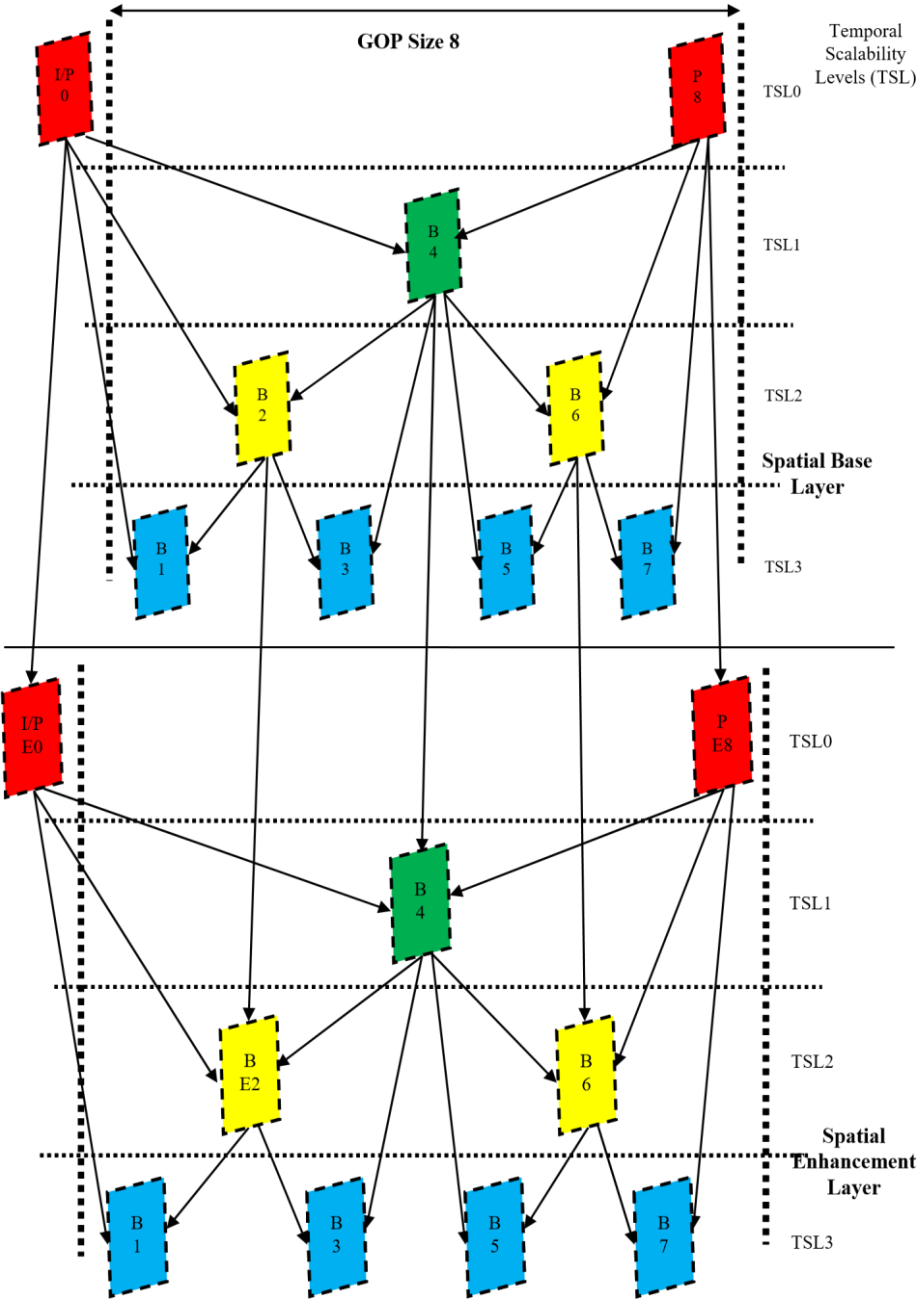

**Figure 1** Temporal scalability levels in a spatial base and a spatial EL (see online version for colours)

The quality scalability is the feature enhancement for pictures of the same spatio-temporal resolution. An adaptive distortion-based intra-rate estimation algorithm has been proposed in Yan and Wang (2009) for improving the video quality of H.264/AVC. An improved context-based adaptive variable length coding scheme has been proposed in Heo and Ho (2010) which modifies the relative entropy coding parts in H264/AVC for achieving lossless intra-coding. Kau and Leng (2015) proposed a simple gradient evaluation method that evaluates the texture orientation inside the coding block to speed up the encoding of H.264/AVC intra-prediction mode. In Thyagharajan and Ramachandran (2006), methods have been suggested to analyse the quality of video streaming created by a sequence of intra-coded and inter-coded frames. These works improve the coding efficiency but increase the computational complexity. The scalabilities should be achieved by balancing both decoder complexity and coding efficiency. The computation complexity in H.264/SVC depends on the methods used to decide the search range and modes for MBs, and also depends on the methods used for motion estimation. Conventional JSVM uses full search method. It calculates the rate distortion cost (RDC) for all possible modes and chooses the mode with minimal RDC as the best mode. But this method consumes more encoder time and increases computation complexity. To reduce encoder time and complexity, various fast mode decision (FMD) algorithms have been proposed for SVC and these are discussed in Section 2. The proposed mode selection algorithm is discussed in Section 3 which reduces the encoder time without sacrificing peak signal to noise ratio (PSNR) or bit rate. Section4 provides the experimental results and followed by conclusion in Section 5.

## 2 Related work

Ri et al. (2009) classify the existing low-complexity algorithms for inter-mode decision as:

- rate distortion (RD) estimation-based algorithms
- rate distortion optimisation-based algorithms
- non-rate distortion optimisation (non-RDO) algorithms

The RD estimation-based algorithms estimate the rate and distortion with quantised coefficients of the discrete cosine transform (DCT) used for coding. The distortion is proportional to the quantisation error (Tu et al., 2006; Ichigaya et al., 2006) and the rate is related to the number and sum of non-zero quantised DCT coefficients. The RDO-based algorithms use statistical relationships between layers and modes to predict the best mode. The non-RDO-based algorithms use the features like texture and edge information to select the optimal mode. Nine prediction modes are used in H.264/AVC for its single layer coding. But H.264/SVC includes seven macro block modes for inter-prediction ($16 \times 6$, $16 \times 8$, $8 \times 16$, $8 \times 8$, $8 \times 4$, $4 \times 8$, and $4 \times 4$), 13 prediction modes for intra-coding and a SKIP mode. Inter-mode decision requires the estimation of MVs for all possible block types for each MB. The optimisation of mode decision will reduce the computational complexity. BL prediction mode and quarter pixel refinement mode have been added for encoding the EL. In Yu et al. (2008), an algorithm has been proposed for mode selection of inter-frame coding based on Lagrangian cost. Computation complexity reduction methods either reduce the number of modes or reduce the search range for motion estimation. In Li et al. (2006), a FMD algorithm for spatial and temporal SVC is presented. In this method, instead of choosing a mode that has minimum RDC, the redundant modes of the EL are minimised based on the relation between the BL and EL. The duplicate modes at higher layer are removed with the mode information available at the lower layer. This algorithm attains good PSNR, but increases the encoding time due to the full search method used for the BL. A layer adaptive mode decision algorithm and a motion search scheme for hierarchical B-frames has been proposed in Lin et al. (2010). In this method the RDC is estimated for different modes in the ELs and mode conditional probabilities are calculated for different temporal layers. Inter-layer prediction is used for EL if the quantisation parameter (QP) is less than 33 otherwise exhaustive searches will be used for mode selection. In this type of adaptive mode selection, the computational complexity for coding the ELs is remarkably reduced, but the bit rate increases by 1% and the average Y-PSNR loss increases by 0.05 db.

A FMD algorithm presented in Yeh et al. (2010) predicts the mode for EL by statistical analysis. By using Bayesian theorem, it confirms whether the mode is best or not and refines the decision by using Markov process. But this method degrades the PSNR and increases the bit rate. The algorithm proposed in Kim et al. (2009) uses the modes of a co-located MB and its neighbouring MBs at the BL to predict the mode of each MB at the EL. The H.264/SVC distributes all hierarchical B-pictures with in two consecutive key pictures at different temporal scalability levels for achieving various temporal resolutions. The inter-prediction methods used for B-pictures increase the computational complexity. To reduce this complexity a fast-inter-prediction mode decision method is proposed in Lee and Kim (2012). In this method, the pixel values of the current block to be encoded are compared with those of a motion compensated reference block using statistical analysis. In our former work (Balaji and Thyagharajan, 2014), a set of modes based on probability is built for BL and the mode selection for EL is obtained using the correlation between the frames. Balaji and Thyagharajan (2015), an FMD algorithm based on likelihood model identifies the prime mode for all types of sequences. Although this algorithm identifies the modes faster, it lags in terms of PSNR and bit rate due to the use of full search method.

In Liu et al. (2019) the MBs are classified into five activity classes using MVs for efficient mode detection. Frame sequences that contain slow motion or uniform motion tend to have more SKIP mode macro blocks (Grecos and Yang, 2005). Few algorithms (Dhanalakshmi et al., 2019; Wang et al., 2019) discussed to enhance the performance of the scalable extension of HEVC (SHVC). In Dhanalakshmi et al. (2019), a superior step search algorithm is introduced to enhance the coding efficiency and PSNR without much increase in the computational complexity. While in Wang et al. (2019), proposed to early terminate the mode decision process in SHVC using the depth level probabilities for code tree unit (CTU). So the analysis shows that no algorithm would search best mode without the loss of either PSNR or bit rate. The algorithms already available reduce the number of modes but do not improve the matching criteria. To improve the matching criteria even for noisy videos and to enhance the mode selection with reduced complexity a novel macro block pre-classification algorithm is proposed. The algorithms proposed in this paper show significant improvement in mode decision, PSNR and bit rate as compared to JSVM and reduce the computation complexity. We have also compared our algorithm with the FMD algorithms already proposed.

## 3 Proposed optimal mode selection algorithm

Generally, to encode a MB called as current MB or target MB, a close match of that MB is searched in a reference frame. A reference frame will be in the previous layer if inter-coded spatial scalability is required or it may be in the previous temporal scalability level (TSL) otherwise it may even be within the same frame in the case of intra-coding. The search can also be either on macro block boundaries or on pixel boundaries. The current MB to be encoded is compared with the reference MB by estimating the sum of squared differences of the corresponding pixels. This sum is called the error or residue. Noise in the video will increase this error. The positive and negative noise will result in cumulative error even if the noise distribution is zero as proposed in our previous work (Balaji and Thyagharajan, 2017). To decimate the effect of noise and reduce the computational complexity, it is proposed to calculate the sum of differences (SOD) between the corresponding pixels as shown in equation (1).

$$SOD = \left| \sum_{i=10}^{15} \sum_{j=0}^{15} [MB_c(i,j) - MB_r(i,j)] \right| \quad (1)$$

where $MB_c(i, j)$ is the current MB to be encoded and $MB_r(i, j)$ is the MB in the reference frame which is to be checked that whether it is a match or not.

The motion of any object is closely related to the center of gravity (COG) of that object. If an MB contains an object, the movement of the object affects the COG of that MB. If the COG of current MB is denoted as $(GX_c, GY_c)$ and COG of the reference MB is denoted as $(GX_r, GY_r)$ then the movement of COG in the horizontal direction $(X_d)$ can be given as shown in equations (4).

$$GX_c = \left(\sum_{i=0}^{15}\sum_{j=0}^{15} MB_c(i,j) \times i\right) \Bigg/ \sum_{i=0}^{15}\sum_{j=0}^{15} MB_c(i,j) \qquad (2)$$

$$GX_r = \left(\sum_{i=0}^{15}\sum_{j=0}^{15} MB_r(i,j) \times i\right) \Bigg/ \left(\sum_{i=0}^{15}\sum_{i=0}^{15} MB_r(i,j)\right) \qquad (3)$$

$$X_d = GX_c - GX_r \qquad (4)$$

The movement of COG in the vertical direction ($Y_d$) is given by equation (7)

$$GY_c = \left(\sum_{i=0}^{15}\sum_{j=0}^{15} MB_c(i,j) \times j\right) \Bigg/ \sum_{i=0}^{15}\sum_{j=0}^{15} MB_c(i,j) \qquad (5)$$

$$GY_r = \left(\sum_{i=0}^{15}\sum_{j=0}^{15} MB_r(i,j) \times j\right) \Bigg/ \sum_{i=0}^{15}\sum_{j=0}^{15} MB_r(i,j) \qquad (6)$$

$$Y_d = GY_c = GY_r \qquad (7)$$

Since the horizontal movement of an object in an MB is represented by $X_d$ and the vertical movement is represented by $Y_d$, the change (or movement) in the COG from the reference MB to target MB can be given as shown in equation (8), where DCOG is the distance from the COG of reference MB to the COG of the target MB.

$$DCOG = \sqrt{(X_d^2 + (Y_d^2))} \qquad (8)$$

This *DCOG* is used to classify the MBs as discussed below.

Table 1     Classification of macroblocks

| MB's class | Nature of the video | Remarks |
| --- | --- | --- |
| C1 | Stationary background OR static foreground | SOD and DCOG are small, SKIP mode |
| C2 | Uniform motion foreground OR smooth motion background | SOD and DCOG values are larger than the previous case (C1) |
| C3 | Non-uniform slow motion background AND foreground | SOD and DCOG are medium |
| C4 | Fast OR complex motion | SOD and DCOG are large |

In adaptive mode selection strategy, a sub set of intra or inter-prediction modes will be sent for the RDO process. Gorur and Amrutur (2014) reduce the bandwidth and computational cost for surveillance video encoders by performing SKIP mode selection using Gaussian mixture model-based segmentation. They also classify the MBs as foreground and background MB and hence reduce the cost of coding of uncovered background regions. In

Yu et al. (2008), skip mode for current MB is chosen when either the co-located macro block in the reference frame is encoded with skip mode or at least one skip mode MB is found above or to the left of the current MB. When skip mode is used, the content of the reference macro block is directly copied to the co-located MB in the current frame and it requires neither motion compensation nor Lagrangian estimation. In our method larger block size is used for coding MBs if the background of a video is stationery and foreground is static. If the foreground has uniform motion with rigid objects or if the background has smooth motion then the residual in encoding will be small (SOD is also small) and hence larger block sizes (16 × 16) are used for encoding. For regions with motions of different objects or generally for regions with complex motions, smaller block sizes are used. If the MB to be encoded belongs to a background scene, there will not be any motion and the MV will be almost zero. In this case SOD and DCOG will be very small and hence SKIP mode is used. If the MB contains slow moving objects, then MV will be of reasonable value and hence all inter modes are used. In this case the SOD and DCOG will be of medium values. If the MB belongs to a fast-moving object, then the MV will be large. Hence the search range should be maximum with all possible inter and intra modes. In this case SOD and DCOG will also be large values. Table 1 summarises these discussions. The mode in an inter-layer prediction is determined by the mode of co-located MB in the frame of BL or previous temporal level. The mode for intra-coding is determined by the modes of the neighbouring MBs in the same frame. First, we are estimating the SOD and DCOG for the collocated MBs in the target and reference frames.

The following algorithm is used to decide the mode selection where $k_1$ and $k_2$ are constants. Since the values of SOD and DCOG depend on the QP used, the constants $k_1$ and $k_2$ are also related with QP in the algorithm. In practice, all video encoders have used a fixed search range in order to obtain uniform quality. But most of the times, the MVs are very small compared to the given search range. The search range for hierarchical B-picture is fixed to 32 in the reference software. Fixed search range has much redundancy. Therefore, in our method we are using adaptive search range. Smaller values of SOD and DCOG indicate that the motion is less. So, the search range is fixed to be small.

The proposed mode selection algorithm defines the size of the search range and the modes to be checked in an adaptive manner. The defined model parameters such as $k_1$ to $k_3$ and $d_1$ to $d_3$ are set (several iterations) to define the search range adaptively. As we know, there are 7 modes to be checked for inter-layer coding and 13 modes for intra-layer coding.

The algorithm works as follows,

The search range will be set as 2, if SOD lies below the product of QP and $k_1$; and DCOG lies below the product of QP and $d_1$. Now the set of modes to be checked will be only 2 modes (SKIP, 16 × 16).

The search range will be set as 4, if SOD lies below the product of QP and $k_2$; and DCOG lies below the product of QP and $d_2$. Now the set of modes to be checked will be only 4 modes (16 × 16, 16 × 8, 8 × 16, 8 × 8).

The search range will be set as 8, if SOD lies below the product of QP and $k_3$; and DCOG lies below the product of QP and d3. Now the set of modes to be checked will be only 7 modes (16 × 16, 16 × 8, 8 × 16, 8 × 8, 8 × 4, 4 × 8, 4 × 4).

The algorithm will check all inter and intra mode, if SOD lies above the product of QP and $k_3$; and DCOG lies above the product of QP and $d_3$ and the algorithm is depicted below.

### 3.1 Mode selection algorithm

Input: $k_1, k_2, k_3$  $k_1 < k_2 < k_3$
       $d_1, d_2, d_3$  $d_1 < d_2 < d_3$

```
MBsizeInterM = {16 × 16, 16 × 8, 8 × 16, 8 × 8, 8 × 4, 4 × 8, 4 × 4}
MBsizeIntraM = {13 modes}
if ((SOD ≤ (QP × k₁)) & (DCOG ≤ (QP × d₁)))
{
   Search range = 2
   mode ∈ (SKIP, 16 ×
   16)
} else if ((SOD≤(QP×k₂)) &
(DCOG≤(QP×d₂)))
{
   Search range = 4 mode ∈ (16 ×
   16, 16 × 8, 8 × 16, 8 × 8)
} else if ((SOD≤(QP×k₃)) &
(DCOG≤(QP×d₃)))
{
   Search range = 8 mode ∈ (16×16, 16×8,
8×16, 8×8, 8×4, 4×8, 4×4) } else
   {
      Search range = 32 mode ∈ (any one of
      the all inter and intra modes)
   }
```

## 4 Results and discussion

The proposed optimal mode selection algorithm is executed using the standard JSVM (joint scalable video model) reference software with version 9.19.15 (Reichel et al., 2007). The testing platform configures to Intel i3 Processor, CPU with 2.40 Ghz, 2 GB RAM. Standard benchmark sequences such as Bus, City, Foreman, Crew, Soccer and Mobile test sequences are taken for the performance evaluation of the proposed algorithm. All the test sequences are set with the parameters as in Table 2.

**Table 2** Simulation parameters

| Parameters | BL | EL1 | EL2 |
| --- | --- | --- | --- |
| Resolution | QCIF | CIF | CIF |
| Frame rate | 15Hz | 15 Hz | 30 Hz |
| QP (BL/EL1/EL2) | 16/20/24, 20/24/30, 24/28/32, 28/32/36 | | |
| No. of frames | 150 | | |
| Maximum delay | 1,200 ms | | |
| GOP size | 16 | | |
| No. of reference frames | 1 | | |
| Search mode | Fast search | | |
| Search range | 32 | | |
| EL search range | 4 | | |
| Fast bidirectional search | On | | |
| Iterative search range | 8 | | |
| Codec | JSVM 9.19.15 | | |

The proposed algorithm was evaluated under three performance measures such as PSNR, bit rate and time. Table 3, shows the experimental results of the six video sequences averaged under different QPs. Each video sequence was iterated with 150 frames to analyse the performance measures for the proposed algorithm. This is compared with the algorithms proposed by Li, Yeh, Kim and Lee.

**Table 3** Average change in BD YPSNR (Bjontegaard, 2001) with respect to JSVM 9.19.15

| Algorithms | Video sequence | | | | | |
| --- | --- | --- | --- | --- | --- | --- |
| | Bus | Foreman | City | Crew | Mobile | Soccer |
| Li | –0.12 | –0.20 | –0.11 | –0.12 | –0.17 | –0.17 |
| Yeh | –0.11 | –0.19 | –0.10 | –0.11 | –0.10 | –0.16 |
| Kim | –0.08 | –0.11 | –0.14 | –0.06 | –0.09 | –0.07 |
| Lee | –0.11 | –0.13 | –0.14 | –0.13 | –0.11 | –0.12 |
| Prop | –0.07 | –0.09 | –0.10 | –0.04 | –0.05 | –0.09 |

Tables 3, 4 and 5 show the comparison among the previously proposed algorithms in terms of BD PSNR, Bit rate and Time. The proposed algorithm achieves 57.44% faster in time with PSNR improvement by 0.43 dB and 0.23 kbps reduction in bit rate compared with the standard JSVM reference software. The results obtained are on an average of all six sequences. All the existing FMD algorithms perform better and save time compared to JSVM at all quantisation values. Although, the proposed algorithm lags behind few existing algorithms in terms of FMD, it outperforms from the previously proposed algorithms in terms of PSNR and bit rate.

Table 3 lists the BD PSNR obtained for the previously proposed algorithms and the mode selection algorithm. It is interesting to note that the proposed algorithm outperforms all algorithms in terms of PSNR. The proposed algorithm provides better visual quality compared to all other algorithms except soccer sequence, and it is due to fast motion with large spatial details.

**Table 4** Average change in BD bit rate (Bjontegaard, 2001) with respect to JSVM 9.19.15

| Algorithms | Video sequence | | | | | |
|---|---|---|---|---|---|---|
| | Bus | Foreman | City | Crew | Mobile | Soccer |
| Li | 2.06 | 3.33 | 2.12 | 2.56 | 2.15 | 2.38 |
| Yeh | 1.06 | 2.32 | 0.84 | 1.44 | 0.92 | 1.23 |
| Kim | 1.59 | 2.46 | 1.79 | 2.55 | 1.79 | 2.45 |
| Lee | 1.44 | 2.80 | 1.75 | 1.93 | 1.68 | 1.53 |
| Prop | 0.82 | 1.28 | 0.59 | 0.63 | 1.12 | 0.68 |

Table 4 lists the BD bit rate obtained for the previously proposed algorithms and the mode selection algorithm. It is interesting to note that the proposed algorithm outperforms all algorithms in terms of bit rate. The proposed algorithm provides better coding efficiency compared to all other algorithms except mobile sequence, and it is due to slow motion convergence with small spatial details.

Table 5 lists the BD encoder time saving for the previously proposed algorithms and the mode selection algorithm. It is more interesting to note that the proposed algorithm outperforms all existing algorithms in terms of saving time for the encoder. The proposed algorithm provides a mode decision algorithm that can able to minimise the complexity of the encoder and achieves fast mode selection.

**Table 5** Average percentage change in computation time, BD Time (Bjontegaard, 2001) with respect to JSVM 9.19.15

| Algorithms | Video sequence | | | | | |
|---|---|---|---|---|---|---|
| | Bus | Foreman | City | Crew | Mobile | Soccer |
| Li | 31.67 | 26.98 | 30.41 | 34.35 | 27.21 | 31.96 |
| Yeh | 34.15 | 31.29 | 32.51 | 38.03 | 31.83 | 35.84 |
| Kim | 37.03 | 31.91 | 31.90 | 36.22 | 35.64 | 36.14 |
| Lee | 34.85 | 31.64 | 33.10 | 39.03 | 31.59 | 35.93 |
| Prop | 38.47 | 31.66 | 34.35 | 36.78 | 41.23 | 39.56 |

In general, the proposed mode selection algorithm provides good time saving for the encoder along with better coding efficiency and good visual quality.

## 5  Conclusions

SVC with its scalable feature appears more flexible with any type of gadgets than advanced video coding (AVC). In addition, the computation is also more than AVC. More FMD algorithms are developed to save encoding time by reducing the complexity in computation but compromising with PSNR and bit rate. The proposed optimal mode selection algorithm based on the orientation of pixels achieved better time saving without any degradation in PSNR or increment in bit rate. The proposed algorithm is simulated with the standard JSVM reference software and found to perform better in all the three measures such as 57.44% time saving, 0.43 dB increments in PSNR and 0.23% compression in bit rate.